\documentclass{PoS}

\usepackage{amsmath}

\title{Top quarks as a probe for heavy new physics}

\ShortTitle{Top quarks as a probe for heavy new physics}

\author{Celine DEGRANDE\\
        Institute for Particle Physics Phenomenology, Department of Physics\\ Durham University, Durham DH1 3LE, United Kingdom\\
        E-mail: \email{celine.degrande@durham.ac.uk}}

\abstract{The heaviest fermion is expected to couple strongly to new physics and appears therefore as a natural probe in many BSM scenarios. Moreover, top physics has now entered in a precision era thanks to the huge amount of top quarks produced at hadron colliders, advanced experimental methods and accurate theoretical predictions. In this talk, we will used effective field theory to search for heavy new physics in a model independent way. This method can also be used to quantify the room left for new physics if no deviation from the SM is found.}

\FullConference{XXII. International Workshop on Deep-Inelastic Scattering and Related Subjects \\
                 28 April - 2 May 2014 \\
                 Warsaw, Poland}

\begin{document}

\section{Introduction}

Effective field theories (EFT) provide a framework to search for new physics too heavy to be produced directly in experiments. Heavy new physics appears then as new interactions between the known Standard Model (SM) particles. As a result, it complements nicely the direct searches for new particles. Those new interactions appear in the Lagrangian as operators of higher dimension built out of the SM fields and invariant under its symmetries,
\begin{equation}
\mathcal{L}=\mathcal{L}_{SM} + \sum_{d>4}\sum_{i} \frac{C_i}{\Lambda^{d-4}} O^d_i,
\end{equation}
where $d$ is the dimension of the operator and $\Lambda$ is the new physics scale. Although this Lagrangian is totally generic, it only becomes predictive when the new physics scale is assumed to be well above the energies probed by the experiments. As a consequence, the ratio of those two scales can be used as an expansion parameter and the main new physics contributions arise from the operators with the lowest dimension, \textit{i.e.} the dimension-six operators in the cases that we are interested in,
\begin{equation}
\mathcal{L}=\mathcal{L}_{SM} + \sum_{i} \frac{C_i}{\Lambda^{2}} O^6_i + \mathcal{O}\label{eq:ld6}\left(\Lambda^{-4}\right).
\end{equation}
Since only a finite set of dimension-six operators can be built, the set of new parameters is now finite and the Lagrangian predictive. 
This only assumption still allows to cover a large class of models, namely all the models with heavy new physics and therefore EFT provide a model independent way to search for new physics.\\
For processes like the top decay to $bW$ or top pair production, the leading contribution to the squared matrix element comes from the interference between the SM and the dimension-six operators,
\begin{equation}
\left|M\right|^2 = \left|M_{SM}\right|^2 + 2 \Re\left(M_{SM}M_{dim6}^*\right) + \mathcal{O}\left(\Lambda^{-4}\right). 
\end{equation}
The contribution at the next order in $1/\Lambda$ cannot be computed from the Lagrangian in Eq.~\eqref{eq:ld6} since they come both from the square of the amplitude with the dimension-six operators and the interference between the SM and the dimension-eight operators. However this contribution is further suppressed and can be neglected.
On the other hand, the square of the amplitude with the dimension-six operators gives the leading contribution for processes where the SM amplitude vanishes or is strongly suppressed like for same sign top pair production,
\begin{equation}
\left|M\right|^2 = \left|0\right|^2 + 2 \Re\left(0 \,M_{dim6}^*\right) + \left|M_{dim6}\right|^2 + 2 \Re\left(0 \,M_{dim8}^*\right)+\mathcal{O}\left(\Lambda^{-6}\right). 
\end{equation}
On the one hand, the new physics contributions are further suppressed in those latter cases. On the other hand, there is no SM contribution from which they have to be distinguished.

\section{Top decay}

\begin{figure}[b] 
\centering
\includegraphics[width=.5\textwidth]{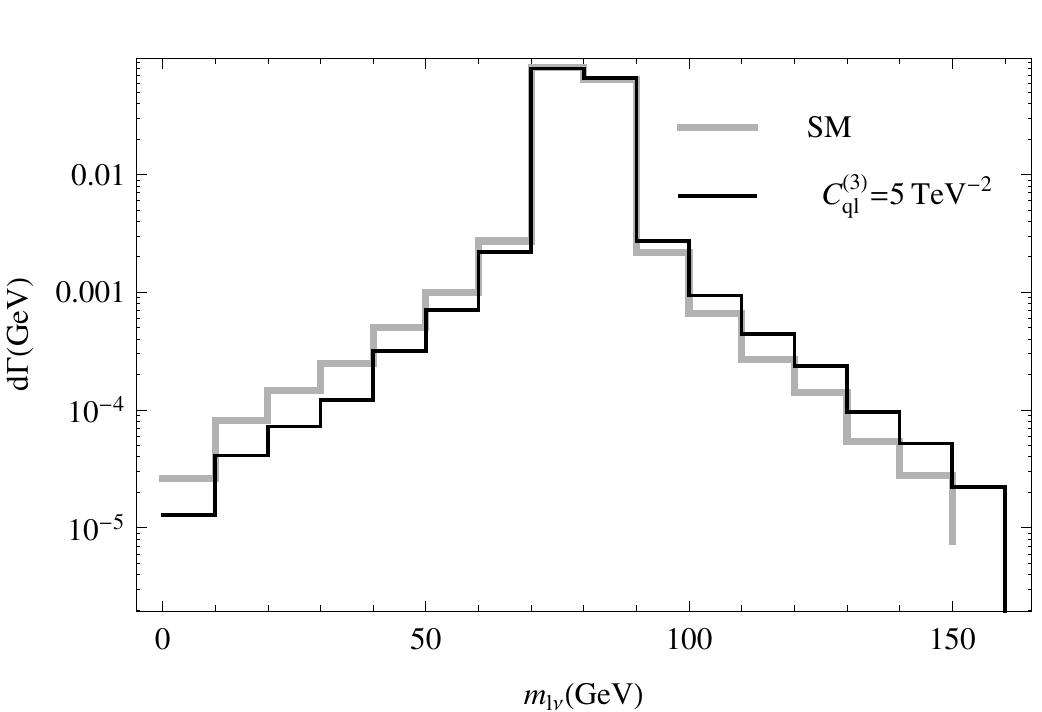} 
\caption{Invariant mass of the lepton and neutrino in the top decay.} \label{fig:mlnu}
\end{figure}

For a massless b quark
%
, only two operators modify the $t \to b W $ decay at the order $\Lambda^{-2}$\cite{Zhang:2010px},
\begin{equation}
\mathcal{O}_{\phi q}^{(3)} = i\left(\phi^\dagger \tau^i D_\mu \phi\right) \left(\bar{q}\gamma^\mu \tau^i q\right)+h.c.\qquad\text{and}\qquad
\mathcal{O}_{tW} = \bar{q} \sigma_{\mu\nu} \tau^i  t \tilde{\phi} W^{\mu\nu}_i, \label{eq:optbw}
\end{equation}
and one operator contributes directly to the $t\to b l\bar\nu_l $ decay\cite{AguilarSaavedra:2010zi} without the exchange of a $W$-boson,
\begin{eqnarray}
\mathcal{O}_{ql}^{(3)} &=& \left(\bar{q}\gamma^\mu \tau^i q\right) \left(\bar{l}\gamma_\mu \tau^i l\right).\label{eq:optbln}
\end{eqnarray}
A similar operator affects to the $t\to b d\bar u $ decay. 
The width is affected mainly by $\mathcal{O}_{\phi q}^{(3)}$ and $\mathcal{O}_{tW}$, 
\begin{equation}
\frac{\Gamma\left(t \to b e^+ \nu_e\right)}{GeV} = 0.1541 + \left[0.019 C_{\phi q}^{(3)}
+ 0.026 C_{tW} + 0 C_{ql}^{(3)}\right]\frac{\text{TeV}^2}{\Lambda^2} \label{eq:width}.
\end{equation}
The angular distribution of top decay products are parametrised by the $W$ helicity fractions,
\begin{equation}
\frac{1}{\Gamma}\frac{d\Gamma}{d\cos\theta}\equiv \frac{3}{8}(1+\cos\theta)^2F_R +\frac{3}{8}(1-\cos\theta)^2F_L +\frac{3}{4}\sin^2\theta F_0 
\end{equation}
where $\theta$ is the angle between the top and the neutrino momenta in the $W$ rest frame.
The helicity fractions are only modified only by $\mathcal{O}_{tW}$,
\begin{eqnarray}
F_0&=&\frac{m_t^2}{m_t^2+2m_W^2}-\frac{4\sqrt{2}{\rm Re}C_{tW}v^2}{\Lambda^2V_{tb}}\frac{m_tm_W(m_t^2-m_W^2)}{(m_t^2+2m_W^2)^2},
\end{eqnarray}
$F_R$ vanishes because the b quark is massless and $F_L$  is then fixed since the sum of the helicity fractions is one by definition. The SM prediction at NNLO \cite{Czarnecki:2010gb} and the CMS measurement  \cite{Chatrchyan:2013jna} constrain
\begin{equation}
 \frac{C_{tW}}{\Lambda^2} = 0.088^{+0.44}_{-0.45} \text{TeV}^{-2} .\label{eq:ctwconst}
\end{equation}
Combining this result with Eq.~\eqref{eq:width}, the CMS result for the top width~\cite{Khachatryan:2014nda} and the SM prediction~\cite{Jezabek:1988iv}, we obtain
\begin{equation}
 \frac{C^{(3)}_{\phi q}}{\Lambda^2} = 0.3^{+1.4}_{-1.2} \text{TeV}^{-2} .
\end{equation}
Finally, the four-fermion operator affects the invariant mass of the lepton-neutrino system in an asymmetric way around the $W$ mass as shown on Fig.~\ref{fig:mlnu}. 
This phenomena explains why its contribution to the total width is nearly zero.

\section{Top pair production}

\begin{figure*}[h!] 
\centering
\includegraphics[width=.58\textwidth]{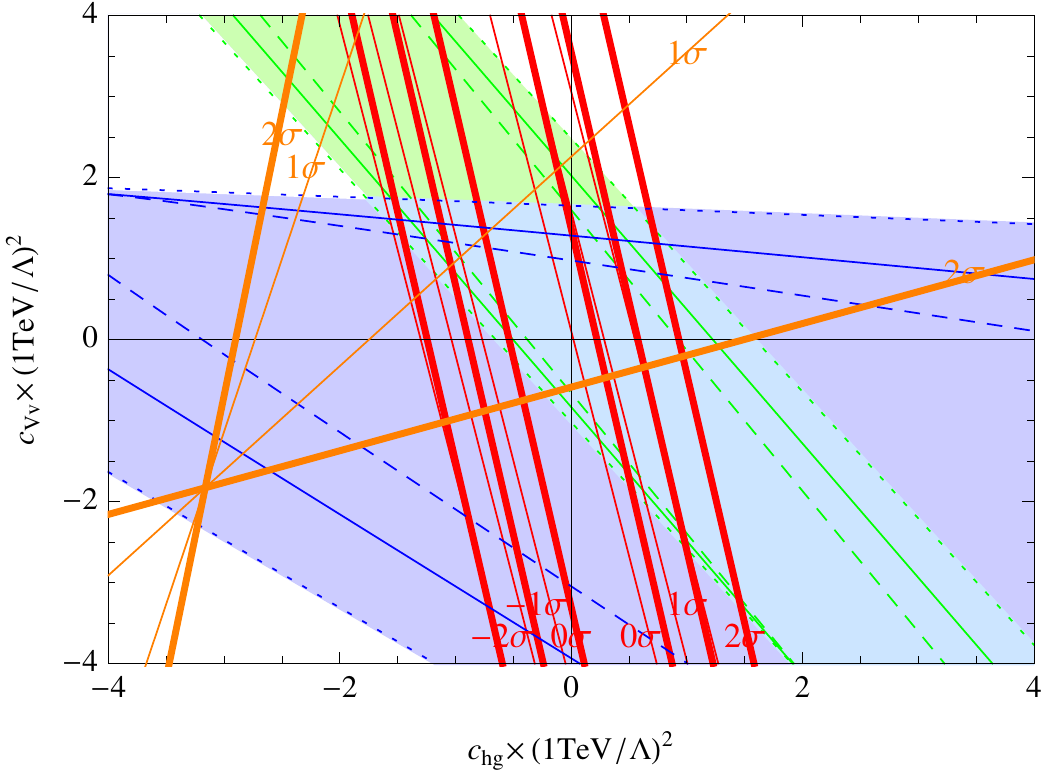} \includegraphics[width=.4\textwidth]{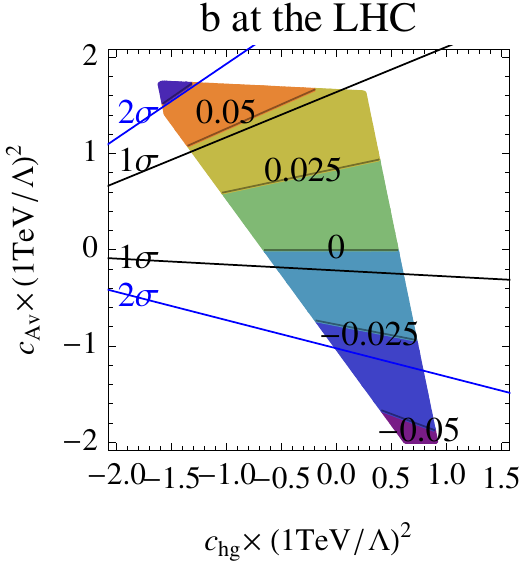} 
\caption{On the left: the constraints from the cross-section measurements at the Tevatron (green), at the LHC at 7 TeV~\cite{Chatrchyan:2012bra} (thin red) and at 8 TeV~\cite{Chatrchyan:2013faa} (thick red), from the invariant mass distribution at the Tevatron (blue) and from the $C$ measurement~\cite{Aad:2013ksa} (orange). On the right: the constraints from the $b$ measurement~\cite{Chatrchyan:2013wua} (blue and black line). The coloured region shows the contribution from the new operators in the area allowed by the cross-section and invariant mass measurements assuming $c_{Av}=c_{Vv}$ ($b=0$ in the SM).} \label{fig:ttbar}
\end{figure*}

Top pair production is mainly affected by five dimension-six operators~\cite{Degrande:2010kt}. Only one operator modifies the interactions of the top and the gluon 
\begin{equation}
 \mathcal{O}_{hg} = \left[ \left( H \bar{Q}_L \right) \sigma^{\mu\nu} T^A t_R \right] G^A_{\mu\nu}\label{chromo}
\end{equation}  
and can therefore have an effect on both gluon fusion and quark annihilation. Only the four-fermion operators with two color-octet currents can interfere with the SM dominant QCD process. 
Two operators involving both the right-handed top and the light quarks 
\begin{eqnarray}
&&\mathcal{O}_{R\, v}  =  \left[ \bar{t}_R \gamma^\mu T^A t_R \right] \sum_{q=u,d}\left[ \bar{q} \gamma_\mu T^A q \right]\qquad\text{and}\qquad
\mathcal{O}_{R\, a}  =  \left[ \bar{t}_R \gamma^\mu T^A  t_R \right] \sum_{q=u,d}\left[ \bar{q} \gamma_{\mu}\gamma_5  T^A q \right]\label{4fermion}
\end{eqnarray}
and two equivalent four-fermion operators with the left-handed doublet of the heavy quarks, named $\mathcal{O}_{L\, v}$ and $\mathcal{O}_{L\, a}$, affect therefore the production by quark annihilation. Consequently, the LHC total cross-section is barely sensitive to the four-fermion operators contrary to the Tevratron one as shown on Fig.~\ref{fig:ttbar}. The total cross-section does not distinguish the left-  and right-handed top quarks and is therefore only sensitive to the combination $c_{Vv}=c_{Rv}+c_{Lv}$. The cross-section measurements give strong constraints also thanks to the improvement the theoretical predictions~\cite{Czakon:2013goa}.
The four-fermion operators have a large effect on the invariant mass distribution since their contribution has an extra $s/\Lambda^2$ factor compared to the SM. On the contrary, the contribution from chromomagnetic operator in Eq.~\eqref{chromo} has to be proportional to the Higgs vev and to the top mass and therefore does not significantly change the invariant mass distribution. Again, this is illustrated on Fig.~\ref{fig:ttbar} by the invariant mass constraint from the Tevatron which depends almost exclusively on $c_{Vv}$. The operators with the axial currents only contribute to odd functions of the scattering angle like the forward-backward asymmetry. In fact, all the asymmetry measurements including the distribution in function of the invariant mass of the top pair or the top rapidity depend on the single combination $c_{Aa}=c_{Ra}-c_{La}$. Nevertheless it gives a pretty good fit of the CDF data~\cite{Degrande:2012zj}. The best fit value for the CDF measurement is still compatible with the LHC results~\cite{Aad:2013cea} which restrict $c_{Aa}/\Lambda^2=-0.72^{+1.77}_{-0.82}$TeV${}^{-2}$. Finally, spin correlation offers an additional way to constrain those operators and in particular to distinguish between operators involving the left- or the right-handed top quark. The distribution of the leptons can be written as 
\begin{equation}
\frac{1}{\sigma}\frac{d\sigma}{d\cos\theta+\cos\theta_-}=\frac{1}{4}\left(1+C \cos\theta_+\cos\theta_-+b_+\cos\theta_++b_-\cos\theta_-\right)
\end{equation}
where $\theta_\pm$ are the angles between the positively (negatively) charged lepton and the top from which it comes from in the top rest frame. The top direction is measured in the top pair centre of mass frame. While the $C$ parameter is sensitive to already constrained parameters,  $b$\footnote{The new physics contribution is the same for both $b$'s.} allows to constrain $c_{Av}=c_{Rv}-c_{Lv}$. Nevertheless, the very good recent measurements of spin correlation provide new competitive constraints on heavy new physics in the top sector as illustrated on Fig.~\ref{fig:ttbar}.

\section{Same sign top pair production}

Out of the five $\Delta F=2$ operators inducing same sign top pair production~\cite{Degrande:2011rt},
\begin{eqnarray}
 \mathcal{O}_{RR} &=& \left[\bar{t}_R\gamma^\mu u_R\right]\left[\bar{t}_R\gamma_\mu u_R\right]\nonumber\\
 \mathcal{O}_{LL}^{(1)} = \left[\bar{Q}_L\gamma^\mu q_L\right]\left[\bar{Q}_L\gamma_\mu q_L\right]\quad&,&\quad
 \mathcal{O}_{LL}^{(3)} = \left[\bar{Q}_L\gamma^\mu \sigma^a q_L\right]\left[\bar{Q}_L\gamma_\mu \sigma^a q_L\right]\nonumber\\
 \mathcal{O}_{LR}^{(1)} = \left[\bar{Q}_L\gamma^\mu q_L\right]\left[\bar{t}_R\gamma_\mu\ u_R\right]\quad&\text{and}&\quad
 \mathcal{O}_{LR}^{(8)} = \left[\bar{Q}_L\gamma^\mu T^A q_L\right]\left[\bar{t}_R\gamma_\mu\ T^A u_R\right],\label{eq:optt}
\end{eqnarray}
only one combination is already constrained by $b$ physics,
\begin{equation}
 \left|c_{LL}^{(1)}+c_{LL}^{(3)}\right|\left(\frac{1\, \text{TeV}}{\Lambda}\right)^2<2.1\times 10^{-4}.\label{UTfit}
\end{equation}
Same sign top pair production is sensitive to the same combination of those operators such that only the three remaining operators are reachable in the near future. The contributions of those operators to the total cross-section are no longer peaked at low invariant mass like for opposite sign top pair production due to the extra $s^2/\Lambda^4$ factor which overcomes the damping from the pdf. Consequently, the whole invariant mass range cannot be trusted when the new physics scale is around the TeV. Nevertheless, one can keep only the region where the EFT makes sense and estimates the expected number of events by putting an upper cut on the invariant mass. For example, the cross-section varies from 500 to 5 fb when $\Lambda$ ranges from 2 to 10 TeV with a cut at $\Lambda/3$ at the LHC at 14TeV.

\section{conclusion}

The top is expected to have new interactions due to its large coupling to the electroweak symmetry breaking sector. EFT provides a model independent to search for heavy new physics: it provides guidance on where deviations from the SM can be found and allows to combine various measurements to constrain the presence of new physics.  The operators involving the top quarks are now been tested at high precision for the first time thanks to the Tevatron and the LHC. The future runs of the LHC will continue to put even more stringent limits on new physics.

\end{document}